\documentclass[12pt,preprint]{aastex}

\newcommand{\simgt}{\lower.5ex\hbox{$\; \buildrel > \over \sim \;$}}
\newcommand{\simlt}{\lower.5ex\hbox{$\; \buildrel < \over \sim \;$}}

\newcommand{\baredth}{\;\overline{\raise1.0pt\hbox{$'$}\hskip-6pt \partial}\;}
\newcommand{\edth}{\;\raise1.0pt\hbox{$'$}\hskip-6pt\partial\;}

\newcommand{\btheta}{\mbox{\boldmath $\theta$}}
\newcommand{\bbeta}{\mbox{\boldmath $\beta$}}
\newcommand{\brc}{\mbox{\boldmath $r_c$}}
\newcommand{\bl}{\mbox{\boldmath $l$}}

\shorttitle{Mass of Galaxy Clusters}
\shortauthors{Wu et al.}

\begin{document}


\title{HOW WELL CAN WEAK LENSING MEASURE THE MASS OF GALAXY CLUSTERS?}

\author{Jun-Mein Wu\altaffilmark{1}, 
Keiichi Umetsu\altaffilmark{2},
Chia-Hung Chien\altaffilmark{1},
Tzihong Chiueh\altaffilmark{1}}
\altaffiltext{1}{Department of Physics, National Taiwan University, Wenshan Chiu, Taipei 116, Taiwan}
\altaffiltext{2}{Institute of Astronomy and Astrophysics, Academia
Sinica,  P.~O. Box 23-141, Taipei 106,  Taiwan, Republic of China}

\begin{abstract}
The technique of weak-lensing aperture mass densitometry, so called the
$\zeta$-statistic, has recently been popular in actual 
observations for measurement of individual cluster mass. 
It has however been anticipated that the
line-of-sight projection by foreground and background matter can 
adversely affect the cluster mass determination with 
not only substantial error dispersion but also 
a sizable positive systematic bias.   
Additionally, the finite number of background galaxies even at a reasonable 
observing depth can also introduce Poisson noise to the mass estimate.
In this paper, we quantitatively investigate the degree of errors separately
contributed by the two sources to the mass determination of 
those galaxy clusters with $M_{200}>10^{14}M_{\odot}$.   
We find that the aperture mass of  $\zeta$-statistic turns out to be 
a mass estimator of much reduced systematic 
bias, due to the cancellation by the positively biased 
local background mass sheet.  
However, the error dispersion of $M_{200}$ arising from 
both projection effect and Poisson noise is found to be still sizable 
($40\%-90\%$),
even for the shear-selected, clean sample 
where multiple clusters located 
within a suitable projected aperture are removed.  
We also investigate how to remedy this large-error problem in weak lensing 
measurements, and propose a plausible alternative mass estimator, 
$M(<\theta_{1000})$, 
an aperture mass
measured within about half the virial radius.  
The aperture mass $M(<\theta_{1000})$ 
is free of bias and has a substantially reduced error dispersion, $39\%$ 
for the worst case of high-$z$, low-mass clusters, 
that can be smaller than the error dispersion of $M_{200}$ 
as much as a factor 3.  
  
\end{abstract}







\section{INTRODUCTION}

The mass function of galaxy clusters has long been recognized as 
the 
most convenient and important indicator for probing the evolution of
structure formation, thereby helping determine the cosmological
parameters.  
In addition, the cluster halo mass, when combined 
with the cluster gas mass via the Sunyaev-Zel'dovich effect, provides 
the opportunity for probing the cluster baryon fraction 
(Umetsu, et al. 2005).  
Conventional techniques, such as measuring the 
velocity dispersion of gravitationally bound galaxies and the X-ray emission
profile, have long been employed to measure the cluster mass, assuming
cluster galaxies and X-ray emitting plasmas 
to be dynamically relaxed within the cluster gravitational potential.  

On the other hand, the new technique of mass measurement through 
weak gravitational lensing has been gaining popularity in recent years, with
the advantage of not having to assume the dynamical equilibrium in the
cluster (e.g., Umetsu, Tada, \& Futamase 1999; Bartelmann \& Schneider 2001;
Schneider 2005). 
This methodology was first pioneered by Tyson et al. (1990).
Various refined techniques 
for the weak-lensing mass determination were
later proposed by several groups (Fahlman et al. 1994; Kaiser 1995;
Bartelmann 1995; Seitz \& Schneider 1996; Squires \& Kaiser 1996;
Broadhurst, Takada, Umetsu et al. 2005).

Among these techniques, the $\zeta-$statistic was particularly devised
to measure the lens mass directly from the tangential component of 
local gravitational image distortions 
without involving a non-local mass reconstruction
(Fahlman et al. 1994; Kaiser 1995).
Schneider (1996) extended the $\zeta$-statistic by generalizing its
kernel, which allows one to define an optimal measure for the detection
of mass concentrations, and this aperture mass 
technique has been applied to deep
optical imaging data to search for clusters
(e.g., Erben et al. 2000; Umetsu \& Futamase 2000; 
Wittman et al. 2001, 2003;
Miyazaki et al. 2002; 
Dahle et al. 2003;
Hetterscheidt et al. 2005;
Schirmer et al. 2006).
King et al. (2001) investigated the cluster mass measurement influenced
by interior substructures, and found the measured mass as accurate as within
$10\%$.  Clowe et al. (2004) studied the effect of asphericity on the
cluster mass determination, and concluded, under the assumption of an
NFW profile, that the asphericality effect generally changes the mass
estimate by $5\%$ to $10\%$.   

In addition, several authors have compared the weak-lensing mass with
the mass determined by the galaxy kinematics and X-ray observations.
Reblinsky \& Bartelmann (1999) concluded that the mass estimates using
$\zeta$-statistic are significantly more accurate than those obtained
from the galaxy kinematics.
Ettori \& Lombardi (2003) studied the mass distribution of 
the rich cluster MS 1008.1-1224 at $z=0.302$ 
based on {\it Chandra} X-ray
and FORS1-VLT multicolor-imaging data, and they
found that the two mass profiles obtained from X-ray and weak-lensing
analyses up to $550 h^{-1}$ kpc
are consistent with each other within $1\sigma$ uncertainty.

Irgens et al. (2002), assuming a singular isothermal sphere model for 
the cluster mass profile, compared spectroscopic velocity dispersions,
$\sigma_p$,
of 13 X-ray luminous clusters around $z \sim 0.3$
with $\sigma_{\rm WL}$ of these clusters
determined by weak-lensing tangential shear measurements out to the 
cluster virial radius.  
It was found that among all, two clusters are in strong discrepancy, 
with $\sigma_{\rm WL} > 2\sigma_p$ and $\sigma_{\rm WL}\approx 2\sigma_p$,
whereas the rest are in fair agreement, 
with $\langle \sigma_{p}/\sigma_{\rm WL}\rangle \approx 1$.
Though these exceptional clusters may be in
dynamical non-equilibrium, another possibility may
arise from the projection effects of 
other mass concentrations
(Cen 1997;  Reblinsky \& Bartelmann 1999;
White, van Waerbeke, \&
mackey 2002; Padmanabhan,
Seljak, \&  Pen 2003; Hamana, Takada, \& Yoshida 2004;
Henawi \&  Spergel 2005)
and/or 
local filamentary structures
(Metzler et al. 1999; Metzler, White, \& Loken 2001)
along the line-of-sight.
Using $N$-body simulations
White et al. (2002) studied the completeness and efficiency
of weak-lensing cluster surveys
on the basis of their {\it mass-selected} mock
cluster sample 
and found that the line-of-sight projection effects can
be quite serious due to the broad lensing kernel.
In the cluster mass estimate based on the convergence map,
they found a positive bias of $\sim 20-30\%$
with a substantially 
larger error dispersion that can even occasionally yield negative 
lens masses.
Metzler et al. (1999; 2001) studied the projection 
effects on weak-lensing mass estimates for massive clusters
caused by the local large-scale filamentary structures.
Including the projection effects from local matter 
within a sphere of $128 h^{-1} {\rm Mpc}$ radius,
they found the lensing convergence maps to yield an positive mass bias 
of $\sim 30\%$ and the mass error dispersion of $\sim  0.3$
for massive clusters at a redshift of $z=0.5$.
In fact, these problems of weak-lensing mass determination, i.e., 
positive mass bias and large mass error dispersion, have been alluded 
in earlier works (Cen 1997; Reblinsky \& Bartelmann 1999).
Further, the cluster halo triaxiality itself can cause a bias in the
lensing-based mass estimation  
(Clowe et al. 2004; Hamana et al. 2004; Oguri et al. 2005),
while it is likely to have less effect on the X-ray cluster mass estimate
(Gavazzi 2005).
Thus, despite that weak lensing offers a unique tool for 
the measurement of
cluster masses without any assumption of their equilibrium state, 
it can however suffer from the projection effects.  
Such problems are less significant in X-ray or spectroscopic
velocity-dispersion measurements.

The present study aims to investigate 
the errors in weak lensing cluster mass measurements
as well.
However, 
this work differs from the aforementioned previous
works, in that we attempt to simulate the actual wide-field 
weak lensing measurements, 
with numerically simulated shear data 
as closely resembling the observing 
data as possible.  
In particular, 
we shall focus on the bias errors and random errors pertinent to 
local weak-lensing measurement of $\zeta$-statistic.
Moreover, at a given observing depth, 
the finite number of background galaxies can
introduce non-negligible Poisson noise convolved with the projection error 
in the measured data.  We shall quantify, in this work,
the regime for which the projection effect dominates, and the other regime 
where the Poisson noise dominates.  

This paper is organized as follows.  
In Sect.~2 we describe our cosmological $N$-body simulations, weak lensing
simulations, and the construction of a mock cluster catalog.
Details of our shear-based mass estimator and mock observations are
presented in Sect.~3.  
In Sect.~4 we apply the shear-based mass estimator to our simulated 
weak lensing observations, and examine 
the statistical properties of the errors in weak lensing cluster mass
estimates.  The radial mass error profiles are discussed in Sect.~5.
Based on the radial error profiles, we propose a plausible alternative 
cluster mass indicator that has much reduced error dispersion.
We finally present the discussions and a summary 
in Sect.~6.

\section{Simulations}

\subsection{Particle $N$-Body Simulation and Weak Lensing Simulation}

We use the Gadget code (Springel et al. 2001) to run 10 sets of
independent simulations with $128^3$ dark matter particles for a 
$\Lambda$CDM model in a $100 h^{-1}$ Mpc cubic box.  
The cosmological parameters are $\Omega_m=0.3$, $\Omega_{\Lambda}=0.7$, 
the Hubble parameter $h = H_0/100 {\rm km/s/Mpc}=0.7$, 
and the linear mass fluctuation amplitude $\sigma_{8}=0.94$.  
The mass resolution is $3.2 \times 10^{10} M_{\odot}$ per particle.  
We also conduct one simulation of 8 times higher mass resolution with 
the same initial condition for checking the convergence of the result.

General methods of backward ray-tracing have been detailed in 
Wambsganss et al. (1998) and Jain, Seljak, \& White (2000).   
In the present study, on the other hand,
we adopt a simpler linear approximation 
for the investigation of weak gravitational lensing.
The shear matrices are generated on
every lens plane with $4096^2$ resolution, and we trace uniform
$1024^2$ rays backwards from the observer plane ($z=0$) 
to the source plane ($z=z_S$).
In the weak lensing approximation that employs the Born approximation, 
each photon travels along the un-perturbed trajectory, 
and the accumulated shear is the linear combination of 
gravitational shear given by every lens plane.    
Throughout this paper, 
we adopt a single source plane located at $z_S=1$,
which is the typical value for the mean redshift of 
background galaxies in an actual weak lensing survey 
with limiting magnitude $R\simeq 25.5$ mag
(see Hamana et al. 2004).  
We then randomly choose a $1 {\rm deg}^2$ field of view from the large 
projected simulation cubic box, 
and generate 36 lensing maps of $1 {\rm deg}^2$ from each simulation.  
Ten sets of independent $100 {\rm Mpc}$ 
cosmological simulations were used to avoid the same clusters to be
observed repetitively.   Hence, the effective survey area is 
$36 \times 10 \times 1 {\rm deg}^2 = 360 {\rm deg}^2$.

To make sure the difference between $128^3$-particle 
simulation and $256^3$-particle simulation 
indeed negligible for cluster mass measurement, 
we compare $256^3$-particle simulation with $128^3$-particle simulation 
of the same initial condition for the weak lensing estimation of cluster
mass.   It is found that two results are almost indistinguishable, 
for the following reason:
The resolution of cluster mass estimation is limited 
by the finite galaxy number density, $n_g \sim  30$ galaxies per arcmin$^2$
which corresponds to the grid resolution of about 
$300\times 300$ per deg$^2$.
On the other hand, even a small cluster of $10^{14}M_{\odot}$ 
of redshift $z=0.2$ contains $3000$ particles within a projected area 
of 10 arcmin$^2$.  
The ten-times higher dark-matter particle number density than the 
background galaxy number density already warrants little 
clumpiness in mass distribution to arise from the discrete particle effect, 
thereby ensuring correct mass estimation even with lower resolution simulations.

\subsection{Mock Cluster Catalog}

We use the friends-of-friends algorithm (Huchra \& Geller 1982) 
with linking-length 0.2 to search for clusters.  
After a cluster is identified, 
the center of mass position $\brc$ is then computed. 
We define the $M_{200}$ of a cluster by increasing the enclosing 
spherical radius 
around the cluster center $\brc$ till it satisfies the criterion 
of $r_{200}$,
within which the mean interior density is 200 times the
mean density $\bar{\rho}(z) =\Omega_m(z) \rho_c(z)$ 
of the universe at the cluster redshift.
Similarly, we define the cluster mass 
enclosed within $r_{\Delta_c}$ by  
\begin{equation}
\label{eq:M200}
M_{\Delta_c} = 
\frac{4\pi}{3}
\Delta_c \bar{\rho}(z) r_{\Delta_c}^3
\end{equation}
with $\Delta_c$ being the mean overdensity within $r_{\Delta_c}$
with respect to $\bar{\rho}(z)$.

In the end, we project the 3-D $\brc$ to obtain
the projected cluster center,
{\mbox{\boldmath $\theta_c$}},  
and the projected $r_{200}$ to $\theta_{200}$, respectively.
The peak position in the 2-D lensing map dose not necessarily coincide 
the center of mass position determined in 3-D particle simulation,
especially in the case where the halo in 2-D projection contains 
substructures or mergers.  From the observational viewpoint,
the cluster center can be chosen from the local surface density peak.  
Nevertheless, we expect the difference not to be so serious as to affect 
the statistics of our mass estimate, since the difference between 
two centers is small compared with the inner most radius where 
the cluster mass estimate is made.  

Through the above procedure, we obtain a mass-selected sample of 
about 4000 clusters with 
$M_{200} > 10^{14}M_{\odot}$ 
between lens redshifts $z=0.2$ to $0.8$ 
in our effective survey area $360 {\rm deg}^2$.
Repeating clusters 
do occur in our  randomly rendered maps but only at 
different cluster orientations and with different 
background and foreground lenses.   
For the purpose of the present work, which
focuses on the measurement errors, repeating clusters can be regarded 
as independent samples since the noise is different.

\section{Method}

Our method for cluster mass estimation
is based on the gravitational shear field,
which is different from the method based on the convergence field
used by Metzler et al. (2001) and  White et al. (2002).
Using the tangential component of galaxy shear
for a lensing mass determination is often a preferred one, 
since it makes direct use of the locally observed galaxy ellipticities
around the target cluster and is widely adopted in recent 
observations (Clowe et al. 2006; Bardeau et al. 2005; Jee et al. 2005).
By contrast, the convergence-based method involves non-local 
mass reconstruction from the observation of galaxy 
ellipticities over a large field.  Mathematically,
the two methods are equivalent, and both exhibit the mass-sheet degeneracy.
But in practice the 
local shear-field measurement has one additional degree of freedom 
in choosing an appropriate
nearby control field as the background mass sheet to be subtracted.

\subsection{Weak Lensing Formalism}

We begin by briefly summarizing 
the basic formalism of weak gravitational lensing.
A general review of weak lensing can be found in
Bartelmann \& Schneider (2001).

The deformation of an infinitesimal ray bundle 
due to gravitational deflection is described by
a mapping between 
the two-dimensional source plane
and 
the image plane as
\begin{equation}
\delta\beta_i={\cal A}_{ij}(\btheta)\delta\theta_j,
\end{equation}
where
$\bbeta$ and $\btheta$
denote the angular position on the source and the image
plane, respectively,
and 
${\cal A}_{ij}$ is the
$2\times 2$ Jacobian matrix of the lens equation.
In the weak lensing regime, the Jacobian matrix
${\cal A}_{ij}$ is symmetric and 
can be decomposed as
\begin{equation}
\label{eq:jacobian}
{\cal A}_{ij}=(1-\kappa)\delta_{ij}-\gamma_{ij},
\end{equation}
where $\gamma_{ij}$ is the shear matrix defined as
$\gamma_{ij}=\gamma_1\sigma_3+\gamma_2\sigma_1$
with $\gamma_i$ being the components of 
complex gravitational shear $\gamma:=\gamma_1+i\gamma_2$,
$\sigma_i$ being the $2\times 2$ Pauli matrices,
and $\kappa$ being the lensing convergence responsible for the
trace-part of the Jacobian matrix.
The lensing convergence $\kappa$ is a 
line-of-sight projection of the matter density contrast 
$\delta=(\rho_m-\bar{\rho})/\bar{\rho}$ out to the source plane $(S)$
weighted by certain combination $g$
of comoving angular-diameter distances (e.g., Jain et al. 2000),
\begin{equation}
\kappa =\frac{3H_0^2\Omega_m}{2c^2}\int_0^{\chi_S}\!d\chi\,g(\chi,\chi_S)
\frac{\delta}{a}=\int\!d\Sigma_m\,\Sigma_{\rm crit}^{-1},
\end{equation}
where $a$ is the cosmic scale factor, 
and $\chi$ is the co-moving distance; $\Sigma_m$ is the matter
surface density $\Sigma_m=\int\!d\chi\,a(\rho_m-\bar{\rho})$ with
respect to the cosmic mean density, and $\Sigma_{\rm crit}$ is the
critical surface mass density of gravitational lensing,
\begin{equation}
\label{eq:sigma_crit}
\Sigma_{\rm crit} = \frac{c^2}{4\pi G}\frac{D_S}{D_L D_{LS}}
\end{equation}
with  $D_L$, $D_S$, and $D_{LS}$ being the angular-diameter distances
from the observer to the lens,
from the observer to the source,
and from the lens to the source, respectively.
The gravitational shear field 
is related with the lensing convergence field
in a non-local manner (e.g., Bartelmann \& Schneider 2001).
The relation between $\kappa$ and $\gamma$ is expressed in Fourier space
as (Kaiser \& Squires 1993)
\begin{eqnarray}
\label{eq:k2g1}
\hat{\gamma}_1(\bl)=\frac{l_1^2-l_2^2}{l_1^2+l_2^2}\hat{\kappa}(\bl),\\
\label{eq:k2g2}
\hat{\gamma}_2(\bl)=\frac{2l_1 l_2}{l_1^2+l_2^2}\hat{\kappa}(\bl),\\
\end{eqnarray}
where $\hat{\kappa}(\bl)$ is the Fourier transform of $\kappa(\btheta)$,
$\hat{\gamma}_i(\bl)$ is the Fourier transform of
$\hat{\gamma}_i(\btheta)$, and $\bl$ is the Fourier variable 
conjugate to the angular position $\btheta$ on the sky.
Equations (\ref{eq:k2g1}) and (\ref{eq:k2g2})
can be used to invert the gravitational shear field to the lensing
convergence field.

\subsection{The $\zeta$- Statistic Mass Estimator}

Jee et al. (2005) performed a weak lensing analysis on a $z=0.8$ cluster, 
and found the mass estimate obtained from the aperture densitometry
(Fahlman et al. 1994),
or so-called the $\zeta$-statistic,
to be very close to that obtained from the
nonlinear iteration-convolution method of Kaiser \& Squires (1993).  
They also investigated the $\zeta-$statistic within 
a projected aperture of $0.15 r_{200}$, where the 
weak lensing assumption almost breaks down, 
and showed only about $10 \%$ mass error.  
Although the $\zeta$-statistic can be accurate
from the linear regime to the weakly nonlinear regime, 
this method however cannot avoid the 
projection effect arising from the 
foreground and background matter along the line-of-sight.  
In what follows, the mass errors caused by projection as
well as other error sources will be quantified.

The observed image distortion of background galaxies can be directly 
used to derive the projected gravitational mass of clusters.
The aperture mass estimate within the angular radius $\theta_1$,
 $M_{\rm \zeta}(<\theta_1)$,
in terms of the tangential component $\gamma_T$ of gravitational shear
can
be expressed as
\begin{equation}
M_{\zeta}(<\theta_1)=\pi (\theta_1 D_L)^2 \Sigma_{\rm crit}\zeta(\theta_1),
\label{eq:zetamass}
\end{equation}
using the $\zeta$-statistic defined as
\begin{equation}
\zeta(\theta_1):=\frac{2}{ 1- \theta_1^2/\theta_2^2}
\int_{\theta_1}^{\theta_2} \! d\ln\theta \, \langle \gamma_T(\theta) \rangle
= \overline{\kappa}(<\theta_1)- \overline{\kappa}(\theta_1,\theta_2),
\label{eq:zetastat}
\end{equation}
where 
$\langle ...\rangle$  denotes the azimuthal average, and
$\overline{\kappa} \equiv 
{\overline{\Sigma}_m}/{\Sigma_{\rm crit}}$ 
is the mean convergence.
Equations (\ref{eq:zetamass}) and (\ref{eq:zetastat})
show that the cluster mass can be measured from 
the galaxies ellipticity within the annulus bounded by $\theta_1$ and 
$\theta_2$ located just outside the mass to be measured. 

As revealed in Eq.~(\ref{eq:zetastat}), the $\zeta-$statistic 
yields the mean convergence interior to $\theta_1$,
 subtracted by the mean background within the annulus 
between $\theta_1$ and $\theta_2$,
$\overline{\kappa}(<\theta_1) - \overline{\kappa}(\theta_1,\theta_2)$. 
Hence, as long as
$\overline{\kappa} (\theta_1,\theta_2) \ll
\overline{\kappa} (<\theta_1)$, the enclosed mass within $\theta_1$
can be obtained by multiplying $\zeta$ by 
the area $\pi \Sigma_{\rm crit} (D_L \theta_1)^2$.  
As the inner radius $\theta_1$ can almost be arbitrarily chosen to obtain the
aperture mass within $\theta_1$, 
when $\theta_1$ is chosen to be $\theta_{200}$, 
the aperture mass $M_{\zeta}$ is approximately the cluster mass,
$M_{200}$.
Obviously, the aperture mass  
is smaller than the enclosed mass 
by a negative compensating mass that serves 
to remove the contribution from a background uniform mass sheet, and
the degree of deviation depends on 
how steep the density profile is.  
In \S 4.3, we will 
demonstrate empirically that the negative compensating mass can 
actually correct 
for the systematic positive bias resulting from the projection effect.
   
The $\zeta$-statistic is a spherically symmetric mass estimator.  
It gives rise to some errors for the typically irregular cluster.  
Nevertheless, this effect has been estimated less than $10\%$ (Clowe et
al. 2004).  
To avoid this effect to influence our cluster mass estimation, the number
of clusters in the sample is crucial.  
Our sample of $4000$ clusters contains
clusters of different sizes and shapes, and the effect of non-spherical lens is
expected to average out.  Moreover, to avoid strong lensing, our mass
estimator avoids the shear measurement near the cluster center (see \S 4.4).   
  
The choice of the parameter $\theta_2$ may also affect the
cluster mass estimate.  For example, 
a small $\theta_2$ will generate large Poisson noise 
since the galaxy number for shear estimate within the annulus bound
by $\theta_1$ and $\theta_2$ is not sufficiently large.  
On the other hand, if $\theta_2$ is too large, 
the cluster mass measurement can be contaminated by the neighboring
clusters in a crowded field.  
In practice, 
we nevertheless obtain similar results for 
$\alpha \equiv {\theta_2}/{\theta_1} =1.15, 1.2, 1.3$ and $1.4$ 
with $\theta_1 =\theta_{200}$.    
Bardeau et al. (2005) and Jee et al. (2005) adopted 
the parameter $\alpha$ of 1.22 and 1.15, respectively.  
Throughout this paper, unless otherwise noted,
we use $\alpha =1.2$ for the measurements of $M_{200}$.

\begin{figure}
\epsscale{1.0}
\plotone{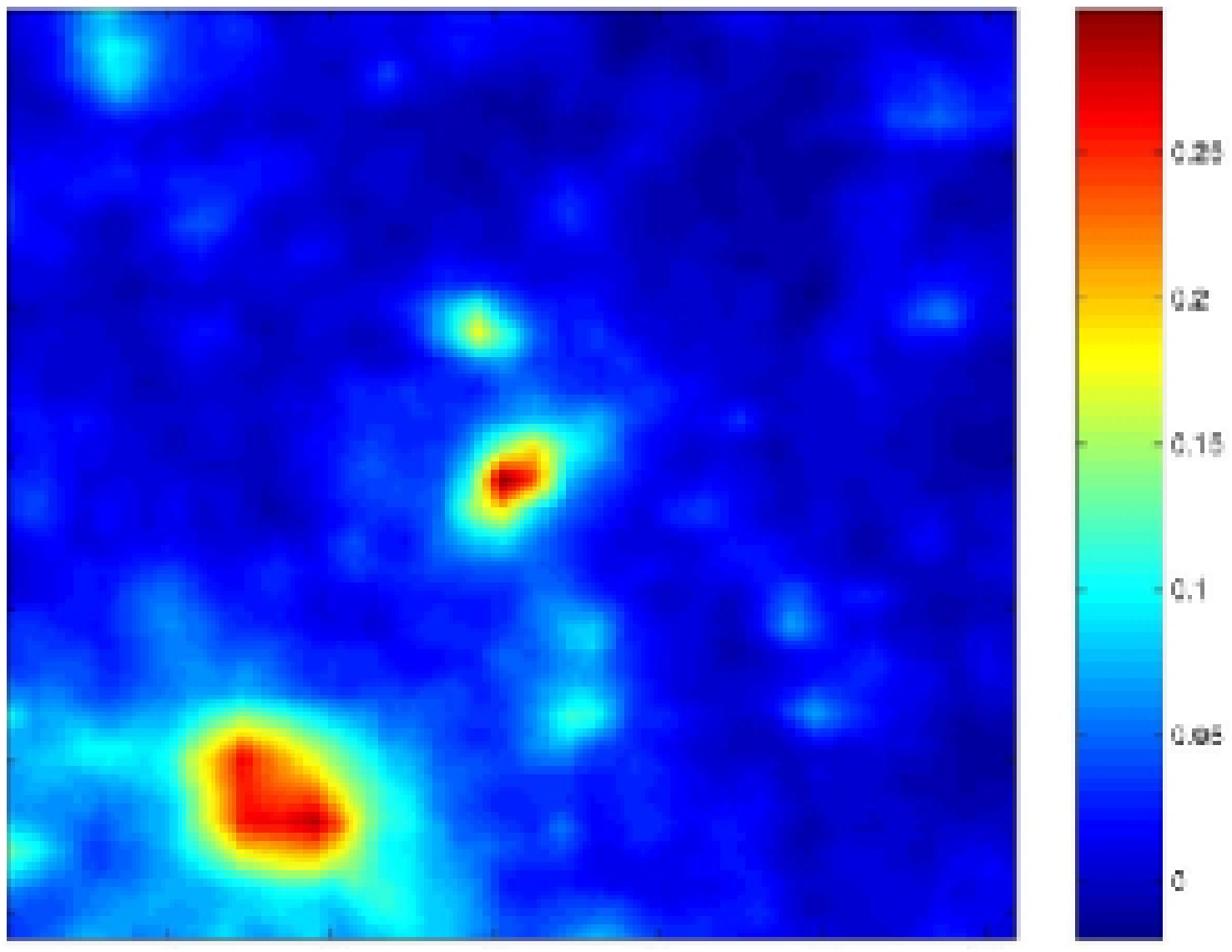}
\caption{An example of the raw $\kappa$ map 
that produces a negative $M_{200}$ 
for the target cluster at the center, 
where the object on the lower-left
corner is located at a different redshift.  
The field of view 
is $1.2 \theta_{200} \times 1.2 \theta_{200}$ defined by the central cluster.}
\label{fig:negativemass}
\end{figure}

\begin{figure}
\epsscale{1.0}
\plotone{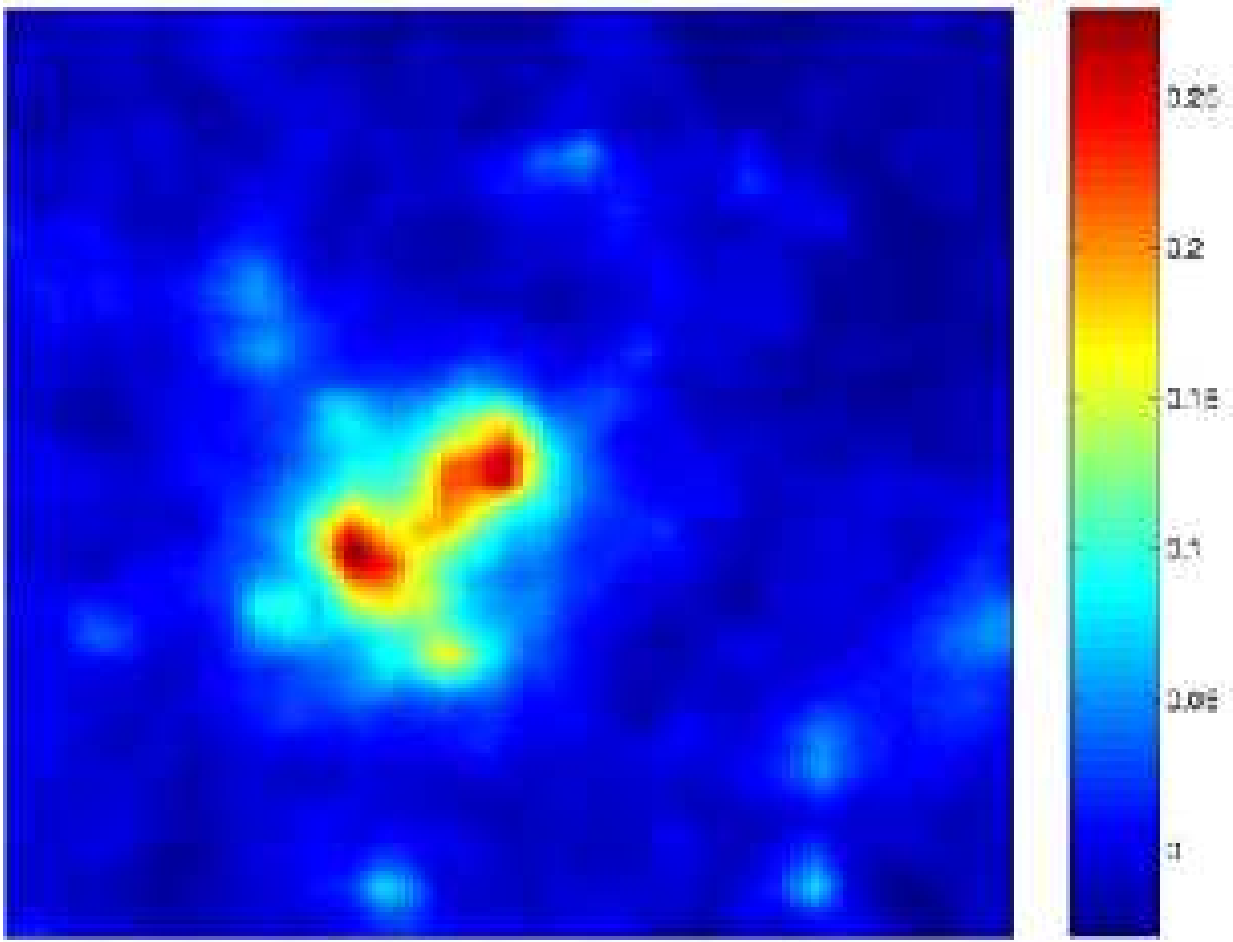}
\caption{
An example of the raw $\kappa$ map that produces an over-estimated 
$M_{200}$,  where the nearby object in the map is also located at a 
different redshift.  The field of view 
is defined in the same manner as in Fig.~\ref{fig:negativemass}.}
\label{fig:overestimatemass}
\end{figure}

For a projected lens system, $\langle \gamma_T \rangle$ 
is produced not only by the cluster itself but also 
by the projected neighboring clusters and large scale structures.
As extreme examples, 
Figs.~\ref{fig:negativemass} and \ref{fig:overestimatemass} 
show two convergence maps that 
give rise to large errors in the cluster mass estimate. 

Figure \ref{fig:negativemass} shows a simulated 
cluster of underestimated $M_{200}$, 
since a high mass cluster of different redshift is located 
just in the annulus bounded by $\theta_1$ and $\theta_2$.  
The underestimated mass can become negative.   
Figure \ref{fig:overestimatemass} shows a cluster of overestimated
$M_{200}$,  
as another cluster of different redshift is enclosed by $\theta_1$. 
In this case the cluster mass is overestimated by a factor $>$2.  
Both Figs.~\ref{fig:negativemass} and \ref{fig:overestimatemass} 
have fields of view of $1.2 \theta_{200}$ defined by the central
objects.  Peculiar clusters of this kind are included in our raw sample of 
4000 clusters.

\subsection{Adding Noise}

Another important factor that strongly influences the mass measurement
is the galaxy intrinsic ellipticity, which plays a role as random noise
in the shear measurement.  In the weak lensing regime, the shear induced
by gravitational lensing is some slight distortion of galaxy images, e.g., 
$10\%$ distortion for cluster lenses.
However, individual galaxies have an intrinsic ellipticity 
of random orientation with considerable dispersion, 
about $30-40\%$.  Such orientation 
noise can only be reduced by taking average over some sufficiently
large projected area around the lens.  

In our simulation, the two components of 
the complex galaxy intrinsic ellipticity,
$\epsilon_{\alpha}^{\rm int}$ $(\alpha=1,2)$,
are generated by Gaussian random numbers
with dispersion $\sigma_{\gamma}/\sqrt{2}$ per component.
In the weak lensing approximation where $\kappa \ll 1$ and $|\gamma|\ll 1$, 
the net ellipticity $\epsilon_{\alpha}^{\rm net}$ is a 
linear combination of galaxy intrinsic ellipticity
$\epsilon_{\alpha}^{\rm int}$
and gravitational shear $\gamma_{\alpha}$, that is, 
\begin{equation}
\label{eq:g2e}
\epsilon_{\alpha}^{\rm net} =  \epsilon_{\alpha}^{\rm int} + \gamma_{\alpha}.
\end{equation} 
In what follows, we assume a weak lensing survey of 
mean galaxy surface number density, $n_g = 30$ arcmin$^{-2}$,
and intrinsic ellipticity dispersion, $\sigma_{\gamma}=0.4$, 
in our simulations.

\subsection{Mass and Redshift Bins of Cluster Sample}

Given a finite number of sample clusters in the simulation, 
we focus on those clusters that can produce sufficiently strong signals
for the weak-lensing mass determination.  
This issue is important in the context of this
study, as we are interested in the observational uncertainties in the
cluster mass determination.  
Clusters of unsuitable mass and redshift
ranges will yield unreasonably high uncertainties to provide useful
astrophysics information.
Our sample of clusters is selected based on
the mass (e.g., White et al. 2002), 
whereas practical weak lensing samples of clusters are
``shear-selected'', that is, weighted by both the mass and the redshift
(e.g., Reblinsky \& Bartelmann 1999; Hamana et al. 2004).
A concise overview of how clusters of viable mass redshift ranges 
are determined is given below. 
It provides a criterion for us to select
those "clean" clusters from 4000 clusters in the raw sample.

Typical cluster search schemes use the convergence map instead of
the shear map (e.g., White et al. 2002; Padmanabhan et al. 2003; 
Hamana et al. 2004).  
In this method one first adopts a Gaussian filter to smooth 
the raw convergence $\kappa$ map, 
followed by identification of local peaks 
in the smoothed convergence map with
a certain threshold.
The threshold signal-to-noise ratio (S/N)
in the smoothed $\kappa$-map
is given  as $\nu=  \kappa_G/\sigma_{\kappa}$
with rms noise in the $\kappa_G$-map,
$\sigma_{\kappa} =  \sigma_{\gamma}/(4\pi \theta_G^2 n_g)^{1/2}$
with the Gaussian smoothing scale $\theta_{G}\equiv {\rm
FWHM}/\sqrt{4\ln(2)}\simeq 0.02 (\sigma_{\gamma}/0.4)(n_g/30 
{\rm arcmin}^{-2})^{-1/2}(\theta_G/1')^{-1}$.

An optimal choice for the threshold
$\nu$ depends on the concentration parameter of the NFW profile, 
$c_{\rm NFW}$, 
the cluster mass $M_{\rm cl}$ and the lens redshift $z_L$.  
Moreover, the optimal threshold $\nu$ can be quantified in terms of the
survey completeness and efficiency.  
For a weak lensing survey with $n_g = 30$ arcmin$^{-2}$ and $z_S=1$,
corresponding to an actual survey
with limiting magnitude $R\simeq 25.5$ mag under a subarcsec seeing
condition, 
Hamana et al. (2004) found that
the threshold $\nu \sim 4$ with $\theta_G \sim 1'$ provides
an optimal balance between survey completeness and efficiency.  
At the optimal threshold $\nu\sim 4$, 
the lower mass detection limits are then 
$M_{\rm cl}=10^{14}M_{\odot}$, 
$2\times10^{14}M_{\odot}$, and 
$4\times10^{14}M_{\odot}$ 
for lens redshift ranges of $0.2<z_L<0.4$, $0.2<z_L<0.6$, 
and $0.2<z_L<0.8$, respectively.    
We therefore divide the cluster mass into three mass bins:
\begin{eqnarray}
10^{14}M_{\odot} &<& M_1 < 2\times 10^{14}M_{\odot},\\
2\times 10^{14}M_{\odot} &<& M_2 < 4\times 10^{14}M_{\odot},\\
M_3 &>& 4\times 10^{14}M_{\odot},
\end{eqnarray}
where $M_1$, $M_2$, and $M_3$ are defined with the $M_{200}$.
The above three redshift bins are also adopted in this study.  
The cluster numbers in different mass bins and different redshift
intervals in our sample are given in Table
\ref{table:rawclusternumber}. 
 
\section{Mass Probability Distribution Function}

For each cluster in our mock cluster sample,
we determine the cluster mass $M_{\rm true}$
using the $N$-body simulation data
and the lensing mass $M_{\rm lens} = M_{\zeta}$
using the $\zeta$-statistic (\ref{eq:zetastat}). 
We note that both $M_{\rm true}$ and $M_{\rm lens}$ are 
projected masses enclosed by a cylinder of certain radius $D_L\theta$
which is relevant to gravitational lensing.
For the true mass $M_{\rm true}$, the integral along the line-of-sight
is taken only within $r_{200}$ from the cluster center, 
while it is taken over the line-of-sight
from $z=0$ to $z=z_S$.
When the angular radius $\theta$ is taken to be 
$\theta_{200}\equiv r_{200}/D_L$, then 
$M(<\theta_{200})\approx M_{200}$.
Collecting all clusters of $M_{200} > 10^{14}M_{\odot}$ 
from the simulation data, 
we then construct the observed mass probability function $P(\mu)$ 
with 
\begin{equation}
\label{eq:massratio}
\mu(\theta) \equiv \frac{M_{\rm lens}(<\theta)}{M_{\rm true}(<\theta)}.
\end{equation}

However, as will be elucidated later, a substantial fraction of lenses
are suffered from the projection effect as well as strong observational
noise due to intrinsic galaxy ellipticities, 
both of which yield considerable errors
in the mass measurement.  
To reduce such errors in the mass determination,
we conduct a
simple-minded "clean sample" procedure, 
which removes those 
clusters from our raw sample that obviously contain nearby
objects in the convergence map, and/or those clusters whose mass is
below a redshift-dependent limiting mass (\S 4.2).

\begin{figure}
\epsscale{1.0}
\plotone{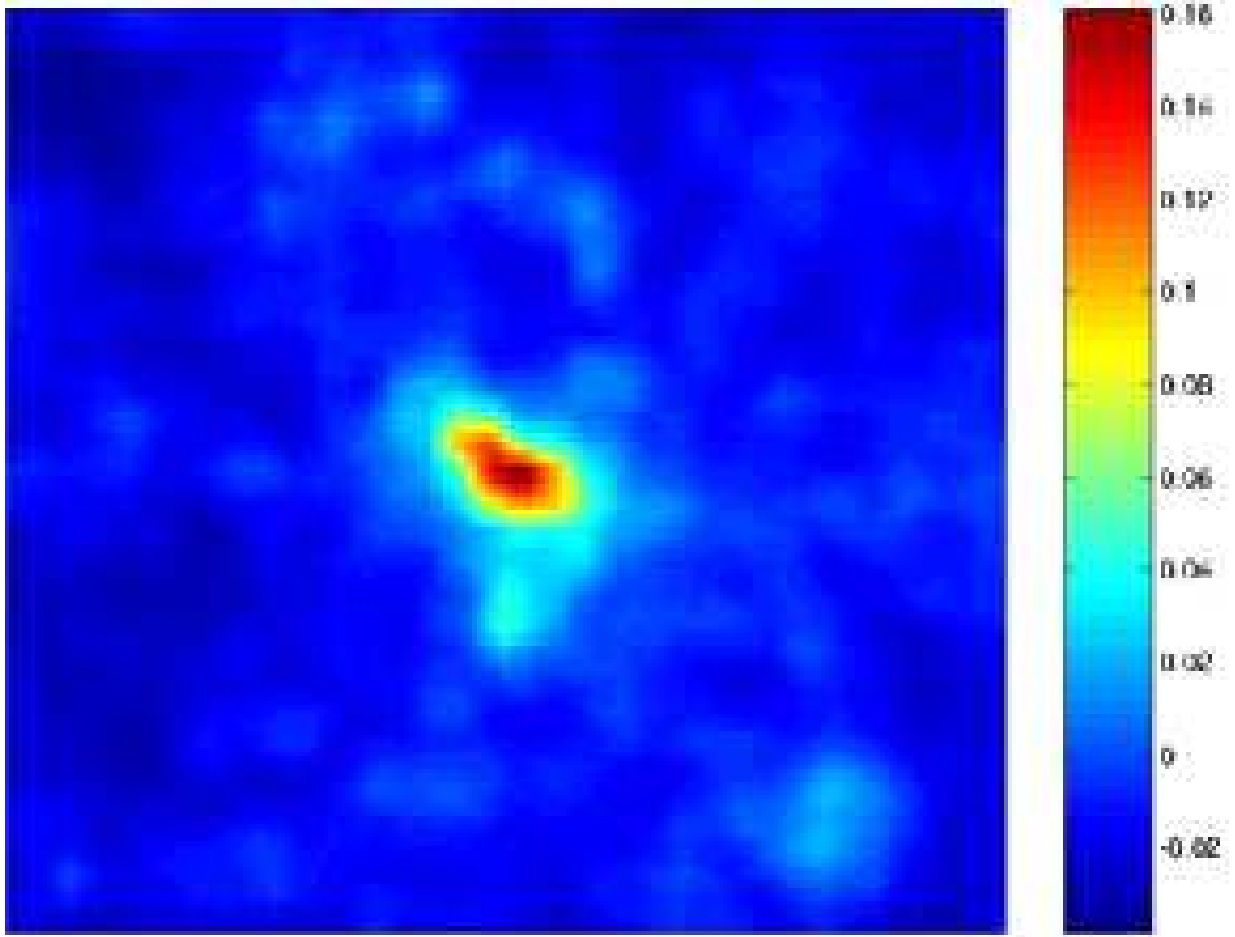}
\caption{An example of clean $\kappa$ maps, 
where contaminants have relatively low signal strengths.}
\label{fig:cleanmap}
\end{figure}

We nevertheless note that even in the clean sample,
mass estimates from the $\zeta$-statistic
still contain residual mass errors due 
to local filamentary structures 
around the target cluster 
as well as intervening large-scale fluctuations (i.e., cosmic shear)
projected 
along the 
line-of-sight, 
as shown in Fig.~\ref{fig:cleanmap}.  
The details of the clean procedure will be presented in Sec.~4.2.  
The numbers of clean clusters in three mass bins and three redshift intervals 
are listed in Table \ref{table:cleanclusternumber}. 
Most of the removed samples contain contaminant objects of different
redshifts due to chance projections along the line-of-sight.
Only a small fraction 
contain physically nearby objects and mergers, 
and the parenthesis in Table \ref{table:cleanclusternumber} gives the
percentage of removed samples that belonging to this category.

\subsection{Raw Sample}

The observational errors of $M_{200}$ is examined
from our raw sample.
The result derived from the raw sample serves as a baseline 
for comparison with those derived from the clean sample.  
The mass probability distribution function is 
defined as $P(\mu)\equiv dn/(N_z d\mu)$, 
where $N_z$ is the total cluster number in a given redshift bin, 
$d\mu$ the mass error bin, 
and $dn$ the cluster number in the mass error bin at a given redshift.
  
\begin{figure}

\epsscale{1.0}
\plotone{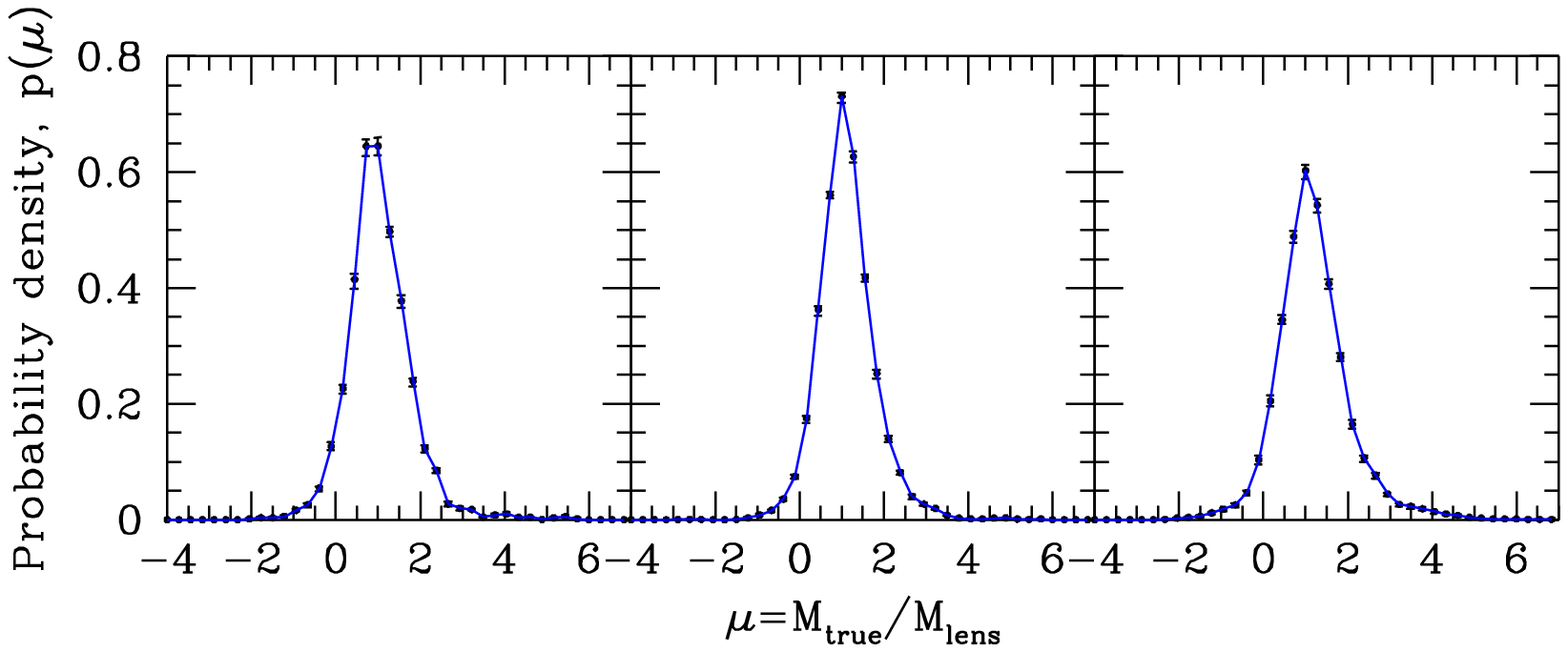}

\caption{
Mass probability distribution functions from noise-free simulations
for three different redshift intervals as indicated in the figure.
The error bars represent 68\% confidence intervals from 20 independent
galaxy realizations.
}
\label{fig:massfreeprobred}
\end{figure}

\begin{figure}
\epsscale{1.0}
\plotone{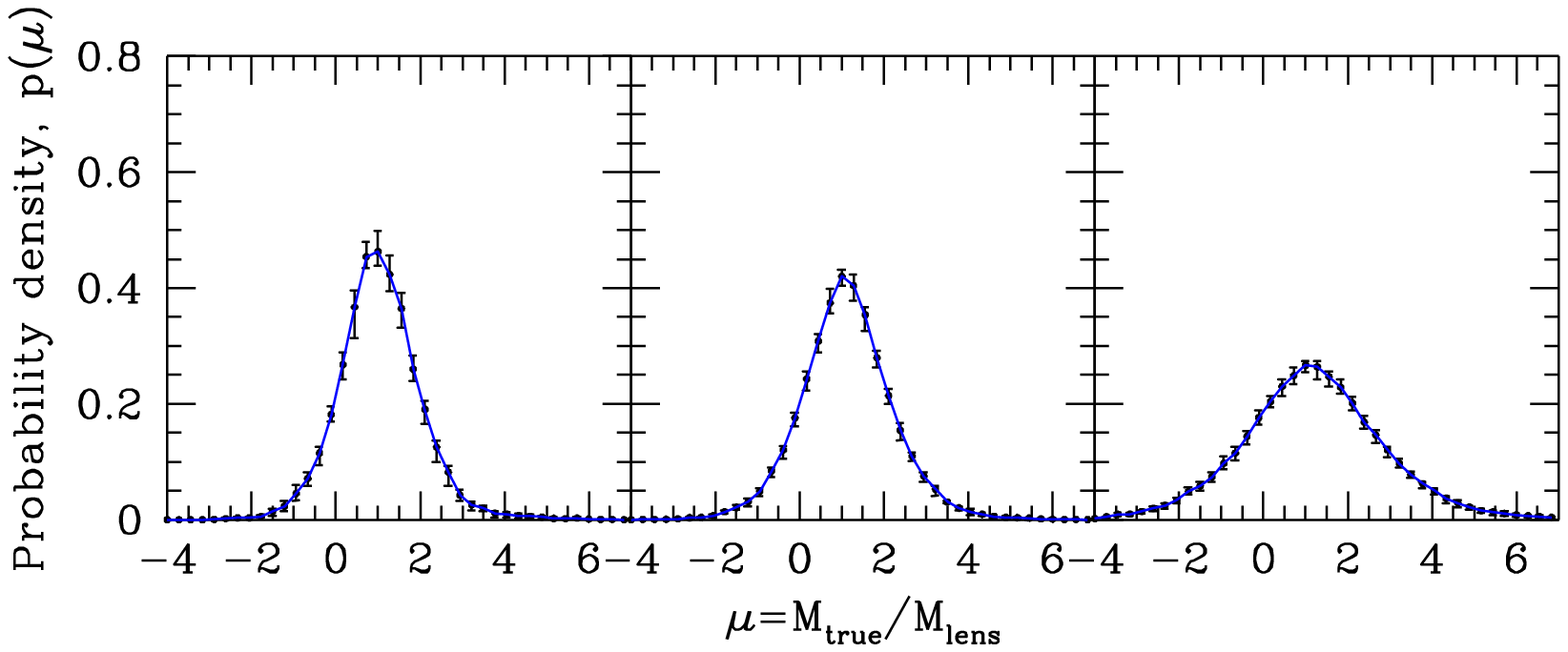}
\caption{
Mass probability distribution functions 
of noisy maps for three different redshift intervals
as indicated in the figure.
The error bars represent 68\% confidence intervals from 20 independent
galaxy realizations.
}
\label{fig:massnoiseprobred}
\end{figure}

Two cases are separately investigated: 
the noise-free case that yields 
Fig.~\ref{fig:massfreeprobred}, 
and the noisy case that yields Fig.~\ref{fig:massnoiseprobred}. 
The noise-free case contains measurement errors 
due solely to the projection effect, 
and the mass probability function is positively skewed with a tail.  
Nevertheless, we find the peak at $\mu$ $\approx 1$, 
which is in agreement with White et al. (2002).  

On the other hand, the noisy case contains contaminations from both 
projection and galaxy ellipticity effects. 
The random Gaussian noise due to intrinsic galaxy ellipticities
adds substantial errors to the
gravitational shear estimate,
giving rise to a mass probability function considerably more symmetric
and broader than the noise-free case.

The error bars in both Figs.~\ref{fig:massfreeprobred} and
\ref{fig:massnoiseprobred} represent the intervals of 
$68 \%$ confidence.
We note that for the noisy case, the confidence intervals are
obtained by averaging over 20 independent galaxy realizations.
Due to the Gaussian nature of the galaxy ellipticity noise, the
S/N can be improved in a predictable manner as $\sqrt{N_g}$, 
where $N_g$ is the number of background galaxies used to determine the cluster
mass.  
Hence, high-redshift clusters, being smaller in angular size, 
have noisier mass estimates than low redshift clusters.  
Furthermore, as high-redshift
clusters are not optimally located for the $z_S=1$ source galaxies, only
massive ones can yield reasonably acceptable mass determination.
   
The mass-error dispersion of the respective mass probability functions
in three different redshift intervals is given in Table
\ref{table:rawsigma}.  
It is clear that the error dispersion for the raw
sample is so large that the mass measurement can hardly be in good use for
construction of the mass function.  
Again, Table \ref{table:rawsigma} serves as the baseline for comparison
against the error dispersion resulting from the clean sample. 

\subsection{Clean Sample}

We apply a "clean" procedure to the raw cluster sample (\S 4.1)
in order to remove clusters that would yield unreliable
mass estimates from the $\zeta$-statistic mass estimator (\S 3.1).
Our clean procedure consists of the following two steps:
The first step is to remove clusters whose masses are below a weak-lensing
detection limit. 
Adopting the result of Hamana et al. (2004),
we first set lower limiting masses of 
$M_{200} = 4\times 10^{14} M_{\odot}$,  
$2\times 10^{14}M_{\odot}$,
and $10^{14} M_{\odot}$,
for clusters in redshift ranges of 
$0.2 < z < 0.8$, $0.2<z<0.6$, and $0.2<z<0.4$, respectively
(see \S 3.3). 
The clean sample contains only these clusters selected 
by the redshift-dependent  limiting masses $M_{200}(z_L)$ above.  
The sample after this first cleaning procedure can be 
regarded as an effective ``shear-selected'' sample of clusters.
The second step is to remove those clusters which 
have detectable "nearby" clusters in the projected map located 
within $1.2\theta_{200}$
from the target cluster.
After the above removal procedure
the remaining randomly rendered maps contain
only the clean clusters that constitute our clean sample.  
Table \ref{table:cleanclusternumber}
shows the cluster numbers of the clean sample in different mass and
redshift bins, as well as the percentage of removed clusters that have
genuinely nearby contaminants in the physical space.

The clean procedure reduces the projection effect in the shear-based
cluster mass estimate, 
and the benefit of "clean" is found to be only moderate as revealed 
by the error dispersions
of the clean sample in Table \ref{table:cleansigma}.  
A comparison between Tables \ref{table:rawsigma} and 
\ref{table:cleansigma} shows that the clean sample has $15\%$ 
improvement for the least massive
clusters of low-redshift to to $50\%$ for
the most massive clusters of high-redshift
in the noise-free error dispersion.  
When observational noise due to intrinsic 
galaxy ellipticities is included, the clean
sample is at most $30\%$ better in error dispersion.  
Such a result cannot be considered to be satisfactory.
Clearly, there is still much room for improvement in error
reduction.  This will be the subject of \S 4.4 and 4.5.

\subsection{Moments of Clean-Sample Mass Probability Distribution Functions}

\begin{figure}
\epsscale{1.0}
\plotone{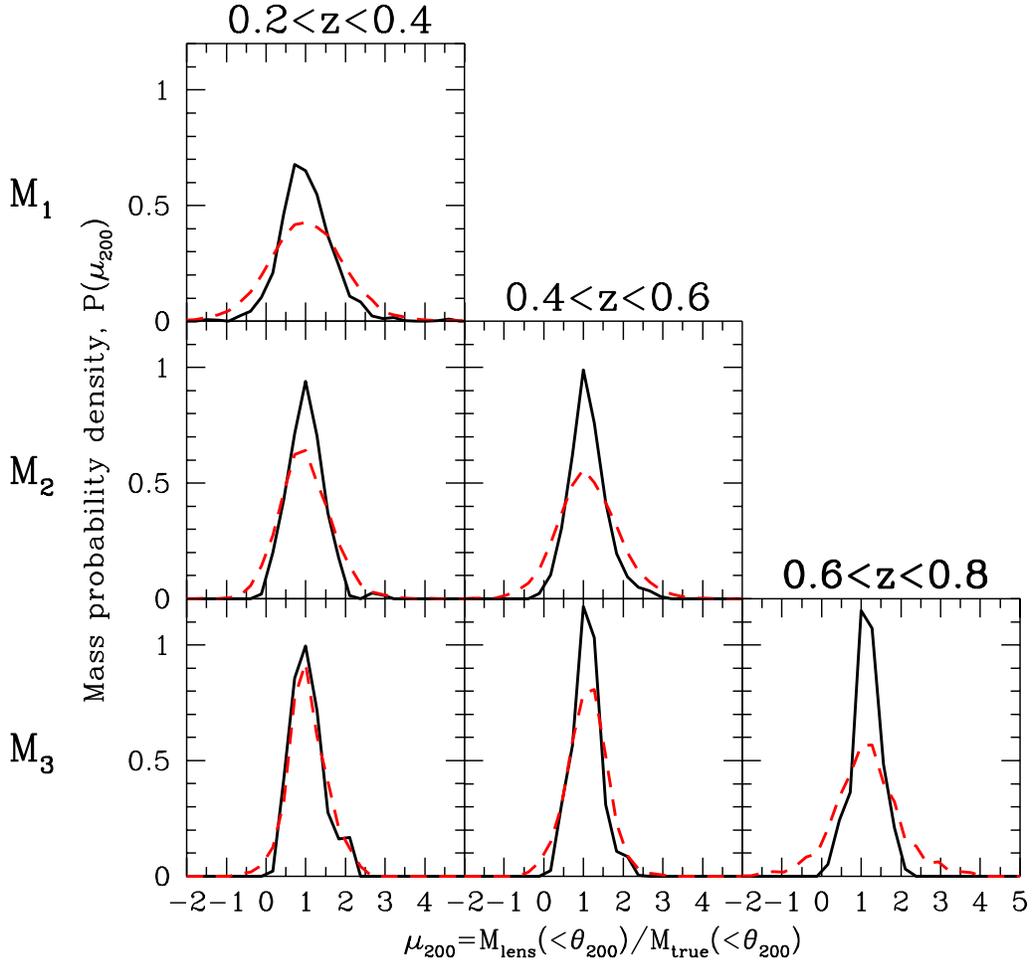}
\caption{
 Noise-free ({\it solid}) and noisy ({\it dashed}) 
 mass probability distribution functions 
 of the clean sample for different bins of 
 cluster mass and redshift: 
 $1 < M_1 /10^{14}M_{\odot} < 2$, 
 $2 < M_2 /10^{14} M_{\odot} < 4$,
 $M_3/10^{14}M_{\odot} > 4$.  Here, the cluster mass
 refers to $M_{200}$.}
\label{fig:massprobdiag}
\end{figure}

Figure \ref{fig:massprobdiag} depicts the mass 
probability distribution functions $P(\mu_{200})$
of $\mu_{200}\equiv M_{\rm true}(<\theta_{200})/M_{\rm lens}(<\theta_{200})$
in noisy and noise-free simulations
for different mass-bins and 
redshift-intervals derived from the clean sample.  
The statistical properties of a probability 
function can be characterized by its distribution moments, 
such as the mean, the dispersion, the skewness, and other higher-order
moments.
We compute these first three moments to quantity the mass probability 
function for the 3 mass bins and 3 redshift intervals.  
The results are listed in 
Tables \ref{table:cleansigma}, 
       \ref{table:cleanmean} 
   and \ref{table:cleanskewness}.

The mass probability function tends to have a positive tail and be
asymmetric.  The positive offset can be characterized by the skewness
$S_3$ tabulated in Table \ref{table:cleanskewness}.  
The skewness is defined
to be the third moment of the distribution function:
$S_3:=\sum_{i=1}^{N}(\mu_i-\bar{\mu})^3/ \left((N-1)\sigma^3\right) $. 
This parameter is a measure of non-Gaussianity originated from the 
projection effect,
since the contaminant lenses are non-Gaussian.  However, inclusion of
the galaxy ellipticity noise substantially reduces the skewness, thus
making the mass probability function more symmetrical about
the peak.  

Note from Table \ref{table:cleanmean} that 
the mean is found to be only slightly 
greater than the peak 
($\approx 1$) shown in Fig.~\ref{fig:massprobdiag} as a result of the small 
positive skewness of the distribution.  
To be specific, the mean has a positive bias of less than $10\%$ for clusters 
of all mass and redshift ranges in our clean sample.  Moreover, the
positive bias increases only slightly with increasing lens redshift.    
The fact that the $\zeta$-statistic mass estimator
has a small bias is surprising, 
since it is at variance with a previous study.   
White et al. (2002) adopted to use the convergence map
for a weak-lensing mass measurement,
and showed a $20\%$ positive mass bias for 
massive clusters of intermediate redshifts.  
Such a positive bias results from
the projection effect by local large-scale structures
surrounding the target galaxy cluster that
accounts for an additional $20\%$--$30\%$ lensing strength 
(Metzler et al. 1999; Metzler et al. 2001).

In contrast to the $\kappa$-based mass measurement, 
the local shear-based mass measurement with $\zeta$-statistic 
turns out to have a compensating  effect to neutralize 
the anticipated positive bias.   
Weak lensing measurements are subject to the mass-sheet degeneracy 
(Bradac et al. 2004), which renders the uniform background matter 
undetectable.   
In the $\kappa$-based mass measurement, 
the overall field tends to be so large as to permit 
an accurate global inversion of the convergence field.
The background so defined is therefore the global background
over a large field.  On the other hand, the shear-based measurement 
is a local measurement and the background matter density is defined
only locally in the projection space.
The second term in Eq.~(\ref{eq:zetastat}) is exactly 
the local background surface density
defined over a control field within an annulus just outside
the target field.   
Hence while the environment surrounding the cluster is rich in matter,
the matter density in the control field just outside the aperture of the
target field can be correspondingly high as to cancel the enhanced 
background contribution along the line-of-sight of the target field.  The
cancellation significantly reduces the positive bias of the lensing mass.

To show the cancellation of the bias, 
we compute the ratio of 
the noise-free local mass 
$\Delta M(\theta_1,\theta_2)=
\pi\Sigma_{\rm crit}D_L^2(\theta_2^2-\theta_1^2)
\bar{\kappa}(\theta_1,\theta_2)$,
contained in the second term in Eq.~(\ref{eq:zetastat}),
to the true cluster mass $M_{\rm true}$,
where all matter outside $50 h^{-1}$Mpc from the target cluster 
has been
excluded for examination of 
the local contribution to the projection effect.
Here we take $\theta_1=\theta_{200}$ and $\theta_2=1.2\theta_1$.
Table \ref{table:annuluskappas} shows
the means and the variances of this ratio for all clusters 
in the clean sample.
The mean ratio ranges from $18\%$ to $26\%$ 
with a considerably small variance.

Indeed, the mean ratio has a correct value to largely cancel the well-known 
mass over-estimation of $20-30\%$ produced by the local projection effect 
(Metzler et al. 1999; Metzler et al. 2001; White et al. 2002).  
This result quantitatively demonstrates
the self-cancellation of two environmental contributions
to be at work with $\zeta$-statistic.  The possibility of such a
cancellation was previously alluded by Metzler et al (1999), 
but the small
number of sample clusters 
prevented them from drawing a quantitative conclusion.


Coming back to Table \ref{table:cleansigma}, we find
the error dispersion for the noise-free case tends to be small 
at redshift intervals in between $0.4< z_L < 0.6$.  
This is because 
the source galaxy redshift $z_S=1$ favors lensing signals from
this lens redshift.  
The noise-free result for the high mass bin, 
$M_3$, at $0.4<z_L<0.6$ can be compared with the result of 
Metzler et al. (2001) for $z=0.5$.  
They obtained an error dispersion $\sigma=0.26$ in the ratio
between the estimated $M_{200}$ and the true $M_{200}$, 
which is in good agreement with 
the noise-free error dispersion in our clean sample,
$\sigma=0.28$.  
On the other hand, for the noisy case, the smallest
error dispersion occurs at lens redshift interval $0.2<z_L <0.4$.  It
means that the errors contributed by galaxy ellipticity noise is
significant, and the larger angular size of low-$z$ lens, having
a larger sampling annulus, helps reduce the sampling noise. 
Table \ref{table:cleansigma} also shows
that the cluster mass measurement errors can be significantly
enhanced by the galaxy ellipticity noise by more than almost $50\%$ for
all lenses, except for the highest mass and lowest redshift bin.  This
indicates that the dominant error source in the weak lensing $M_{200}$
can be the galaxy ellipticity.   This finding motivates us to proceed
on seeking an alternative mass estimator below.

\subsection{Radial Mass Error Profile}

\begin{figure}
\epsscale{1.0}
\plotone{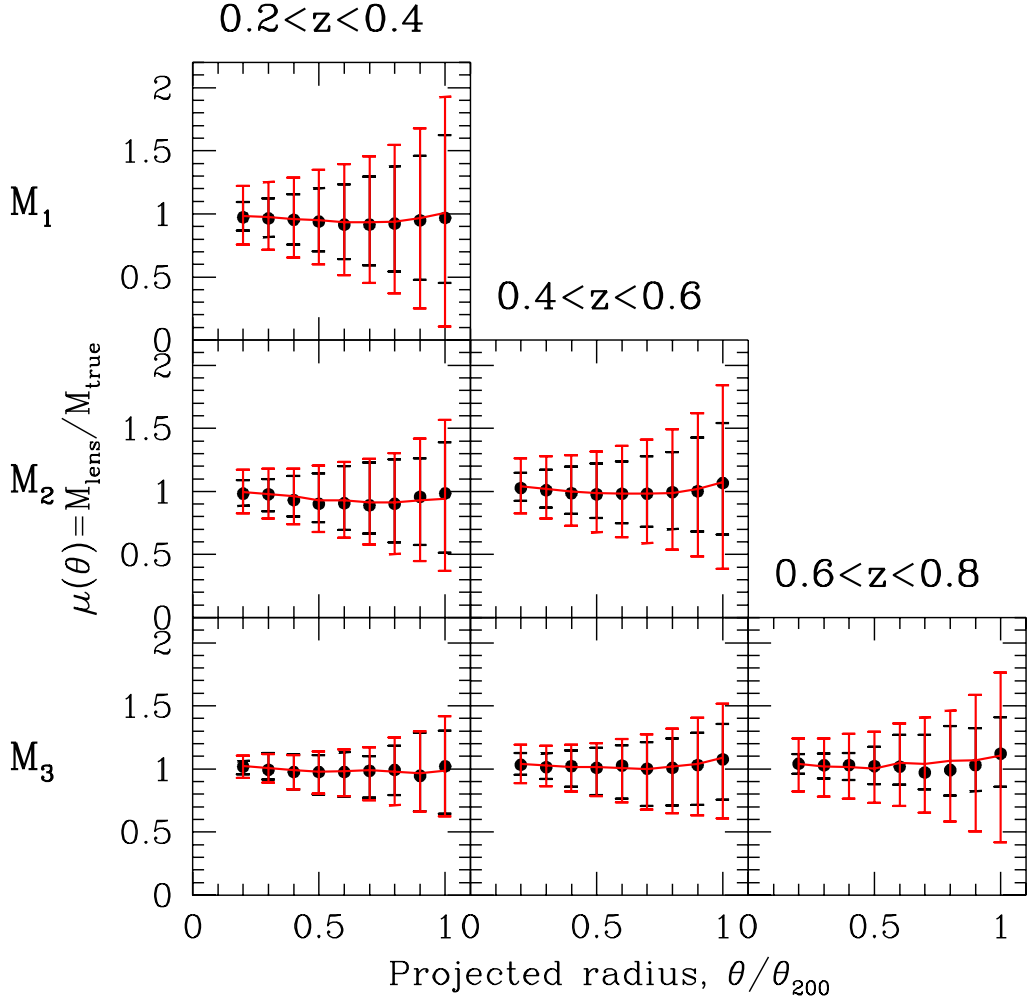}
\caption{
Noise-free ({\it black})
and noisy ({\it red})
projected-mass radial profiles of the clean cluster sample 
for different mass bins and redshift intervals:
 $1 < M_1 /10^{14}M_{\odot} < 2$, 
 $2 < M_2 /10^{14} M_{\odot} < 4$,
 $M_3/10^{14}M_{\odot} > 4$. 
Mean values are represented by filled circles ({\it black})
for the noise-free case,
and by connected lines ({\it red}) for the noisy case.
The error bars indicate 68\% confidence levels 
around the mean.
}
\label{fig:massratiodiag}
\end{figure}

We turn to the radial profile 
$\mu (\theta) = M_{\rm lens}(<\theta)/ M_{\rm true}(<\theta)$,
measured for individual clusters of the clean sample.
We fix the outer radius $\theta_2$ to be $1.2\theta_{200}$
and vary the mass-aperture radius $\theta_1$.  
The results are presented in Fig.~\ref{fig:massratiodiag}
for the noise-free and noisy cases. 
The innermost radius is chosen to be $0.2 \theta_{200}$ 
so as to stay in the observationally
weak-lensing regime.  
For a typical cluster in our clean
sample, $r_{200}$ $\approx 1 h^{-1} {\rm Mpc}$ 
and the innermost radius limit is
about $200 h^{-1} {\rm kpc}$, which is mostly outside the scaling 
radius of the NFW profile.

The error bars in Fig.~\ref{fig:massratiodiag} 
represent $68\%$ confidence levels 
around the estimated mean $\mu(\theta)$,
obtained over 20 independent background realizations
similar to Fig.~\ref{fig:massnoiseprobred}.
The deviation of the mean-$\mu$ from unity indicates
a systematic bias.  
Very little systematic bias is detected interior to
$\theta_{200}$ in Fig.~\ref{fig:massratiodiag}.  

The error bars in Fig.~\ref{fig:massratiodiag}
are seen to decrease with 
decreasing radius.  
In particular,
the mass error caused by intrinsic galaxy ellipticities 
(i.e., the difference between error bars of noisy and noise-free cases)
decreases mostly
noticeably for $\theta_{200}>\theta>0.7 \theta_{200}$.  This feature arises
from the peculiar feature of $\zeta$-statistic, where the enclosed mass 
within a smaller radius is measured by a larger annulus.  
Such an error stays
relatively constant for $\theta \simlt 0.7\theta_{200}$, which can be
understood from the geometry.  The annulus area enhancement begins to
saturate when the inner radius becomes smaller than half the outer
radius.  
On the other hand, the projection errors continue to decrease at the ever
decreasing radius.  
This trend can be understood from the second equality of 
Eq.~(\ref{eq:zetastat}).  
For a small
$\theta_1$, the interior convergence $\overline{\kappa}(\theta_1)$ is
much greater than the average convergence in the annulus
$\overline{\kappa}(\theta_1,\theta_2)$, which contains lensing
contributions of other redshifts.  It therefore becomes increasingly
insensitive to most contaminant lenses for a decreasing $\theta_1$.

\subsection{Alternative Weak-Lensing Mass Estimators?}

Section 4.3 already reveals the considerably large errors arising from
galaxy ellipticity in the weak lensing mass measurements.  On the other
hand, Sec.~4.4 also shows that despite a sizable error in the measured 
$M_{200}$, the error becomes substantially
small for the fractional mass well within $r_{200}$.  This tendency
has also been noticed by Metzler et al. (2001) and Cen (1997) previously
for noise-free mass estimations.

In order for the weak-lensing measurement to yield 
a reliable cluster mass estimate, 
we seek an alternative mass estimator that is contaminated by 
the ellipticity noise and the projection effect 
to a much lesser degree.  
In Table \ref{table:cleansigma500}, 
we enlist the ever-decreasing mass errors 
for the 2D projected mass measured over the 
decreasing radii, 
$\theta_{500}$, $\theta_{1000}$, and $\theta_{1500}$.
Here $\theta_{\Delta_c}\equiv r_{\Delta_c}/D_L$ 
is the projected angular radius 
of the 3D interior mass $M_{\Delta_c}$
defined by Eq.~(\ref{eq:M200}).   
These radii correspond, on average, to $\theta=0.7 \theta_{200}$,
$0.5 \theta_{200}$, and $0.4 \theta_{200}$, respectively, in weak
dependence on the lens mass and redshift.  
Projected masses evaluated at these radii
contain significantly smaller errors than $M_{200}$, 
and may serve as alternative mass indicators.

A similar proposal was put forth for the mass estimator of
X-ray clusters.  With cosmological simulations, 
Evrard et al. (1996)  found that X-ray mass
estimates are remarkably accurate when evaluated 
at radii from $r_{500}$ to $r_{2500}$.  
Recent X-ray observations measure the cluster mass at
$r_{500}$ (Finoguenov et al. 2002) and at $r_{2500}$ (Allen et
al. 2002).   
Moreover, Jee et al. (2005a) compared the cluster mass 
profiles derived from X-ray and from weak lensing, and 
found increasing agreement with decreasing radii.  
The radius $r_{2500}$ is about $0.3 r_{200}$ for our sample clusters.  
In fact, the mass within $0.3 \theta_{200}$ 
may be accurately measured by weak lensing even at a lens redshift 
of $z_L \sim 0.8$, 
as discussed by Jee et al. (2005b).   
However, 
the cluster mass within $0.3 r_{200}$ can be sensitive to 
the complex interior structures of a cluster, for example,
containing two cores in a cluster recently undergoing a merger,
and/or the baryonic physical processes such as radiative cooling 
and galaxy formation.
Moreover, for a dual-core cluster, 
$r_{2500}$ can be ambiguous to define.

We therefore suggest that $M(<\theta_{1000})$ be a better weak-lensing mass
estimator. 
On one hand, it has an acceptable worst mass 
errors ($<39\%$) for
the 3 mass bins and different redshift intervals, as revealed in Table
\ref{table:cleansigma500}.  On the other hand, the radius is still
sufficiently large, $\sim 0.5\theta_{200}$, 
so that the mass estimator can be insensitive to the
cluster interior structures.  
Finally, $M(<\theta_{1000})$ is a significant
fraction of $M_{200}$ and it should still follow the well-known
similarity scaling of the CDM gravitational collapse.

\section{CONCLUSIONS}

This paper reports a systematic study on the accuracy of cluster mass
measurements through weak lensing observations. 
Specifically this work
takes into account the mass errors introduced by the projection effect
and by the Poisson noise of finite number of randomly oriented
background galaxies.  
Among these sources of mass errors, the projection effect
has been reported to yield a non-negligible systematic positive bias 
($\sim 20\%$) in the cluster mass estimate based on weak lensing.
Adopting the local shear-field measurement using $\zeta$-statistic, 
which is often used in actual observations, we nevertheless 
found that such a positive bias can be largely canceled by 
the positively-biased
local background mass sheet.  
That is, 
the $\zeta$-statistic measurement can provide a bias-free 
cluster mass estimate.  

In this paper, we also report that the error in $M_{200}$ determination
is expected to exceed $50\%$ 
even for a moderately deep observation
($R\simeq 25.5$ mag), regardless of lens mass and redshift.  
Even after a clean
procedure that removes clusters with detectable companions in the
projected map, the mass error is still substantial, exceeding $40\%$ for
a $R\simeq 25.5$ observation.
Mass errors can scatter observed data of one mass bin into another bin when
constructing the cluster mass function, and smear out the mass function.  
In the mass range where the mass function has a large
gradient, i.e., $M > M_{\star}$, such mass errors can greatly distort
the mass function.  
The measurement error will eventually propagate into the
determination of cosmological parameters, such as the matter density
parameter $\Omega_m$ and the matter fluctuation amplitude $\sigma_8$, which
rely critically on an accurate mass function. 
 The so-called self-calibration
was devised to correct for the systematic errors of this kind (Hu 2003),
but the random errors are un-removable even with the self-calibration.  

To significantly reduce the mass error in weak lensing measurements,
we also suggest the possibility of an alternative lens mass, which has a
considerably smaller error.  
It is an interior mass well inside $\theta_{200}$.  
For example, 
Fig.~\ref{fig:massratiodiag}
shows that the error in $M(<\theta_{500})$
(mass interior to approximately $0.7 \theta_{200}$) is already noticeably
smaller than $M_{200}$, the error dispersion of
 $M(<\theta_{1000})$ is at most  $39\%$ for
all detectable clusters and the error dispersion of $M(<\theta_{1500})$ 
at most $32\%$.   
Comparing 
Table 
\ref{table:cleansigma500}
with
Table \ref{table:cleansigma},
we find $M_{1000}$  to have more than a
factor 2 in error reduction for all detectable clusters; for low-$z$
lenses, the error reduction can be as large as a factor 3.  Given these
results, we suggest that $M(<\theta_{1000})$ be a better mass variable
than $M_{200}$ for constructing the mass function.

Concerning the mass function of $M(<\theta_{1000})$, 
it should be reminded that the mass function of $M_{200}$ 
cannot be analytically derived,
and needs to be determined empirically from $N$-body simulations.  
From this spirit,
the mass function of $M(<\theta_{1000})$ 
can also be obtained from $N$-body
simulations, in a similar manner as the mass function of $M_{200}$.  
To the best of our knowledge, there have not been investigations on the
mass function for mass different from $M_{200}$ in literature.  Such a
new class of mass function, containing less mass scatter than the
conventional mass function, may be more useful for constraining the
cosmological parameters.
However, whether or not the more accurately measured $M(<\theta_{1000})$ 
function
can actually be more useful than the less accurately measured $M_{200}$
function really depends on the detailed form of the $M(<\theta_{1000})$ 
function
that contains the cosmology-parameter-sensitive feature, similar to a break
at $M_{\star}$ for the $M_{200}$ function.  Therefore our suggestion at
this point in favor of an alternative mass estimator for a new mass
function should be regarded as plausible but still preliminary.

Finally,
though the depth of observation has been fixed to $R\simeq 25.5$ mag 
in this work for which $n_g \simeq  30$ arcmin$^{-2}$, 
for evaluation of mass errors, we may 
relax this constraint to assess other observing depths
straightforwardly.  
At a medium depth $R\simeq 24.5$ mag,
suitable for wide-field surveys, the background 
galaxy density $n_g \simeq 20$ arcmin$^{-2}$
(Fontana et al. 2000).  
One can quickly estimate from
Fig.~\ref{fig:massratiodiag}
with the Poisson statistics that the galaxy ellipticity noise is
still much less than the errors at
$\theta_{500}$, $\theta_{1000}$, and $\theta_{1500}$ 
introduced by the projection effect for clusters of
$M>2\times10^{14}M_{\odot}$ and $0.2<z_{L}<0.4$ and clusters of
$M>4\times10^{14}M_{\odot}$ and $0.2<z_{L}<0.6$.  
That is, a medium-depth observation 
can provide as an accurate mass measurement as
a deep observation 
for those more massive and lower redshift clusters.  
Such information is crucial for the planning of wide-field surveys. 


\section*{Acknowledgments}

The authors would like to thank Masahiro Takada for useful discussions.
This work is supported in part by the grant,
NSC-94-2112-M-002-026, from National Science Council of Taiwan

\clearpage

\begin{table} 
\caption{Cluster number in the raw sample}
\begin{tabular}{@{}lccccc@{}}
\hline\hline
            &   $0.2<z<0.4$    &$0.4<z<0.6$        & $0.6<z<0.8$    \\ 
\hline
$1 < M_{200}/10^{14}M_{\odot}< 2$  &  444 & 1144  & 1374\\
$2 < M_{200}/10^{14}M_{\odot}< 4$  &  145& 350 & 364\\
$ M_{200}/10^{14}M_{\odot} > 4$    &  59  & 92   & 43  \\ 
\hline
\end{tabular}
\label{table:rawclusternumber}
\end{table}

\begin{table}
\caption{Cluster number in the clean sample and percentage 
of interacting clusters among those that are removed during the clean}
\begin{tabular}{@{}lccccc@{}}
\hline
\hline
            &   $0.2<z<0.4$    &$0.4<z<0.6$        & $0.6<z<0.8$    \\ 
\hline
$1 < M_{200}/10^{14}M_{\odot}< 2$  & 324 ($15\%$) &   & \\
$2 < M_{200}/10^{14}M_{\odot}< 4$  &  91 ($20\%$)& 293 ($22\%$)& \\
$ M_{200}/10^{14}M_{\odot} > 4$ &  22 ($16\%$)  & 72 ($18\%$)  & 38
 ($25\%$) \\ 
\hline
\end{tabular}
\label{table:cleanclusternumber}
\end{table}

\begin{table}
\caption{
Dispersion of 
$\mu_{200}=M_{\rm lens}(<\theta_{200})/M_{\rm true}(<\theta_{200})$ 
in the raw sample
}
\begin{tabular}{@{}lccccc@{}}
\hline\hline
            &   $0.2<z<0.4$    & $0.4<z<0.6$       & $0.6<z<0.8$\\ 
\hline
$1 < M_{200}/10^{14}M_{\odot}< 2$
 &  $(0.69, 0.98)$  & $(0.63, 1.13)$  & $(0.74, 1.74)$\\
$2 < M_{200}/10^{14}M_{\odot}< 4$
 &  $(0.55, 0.72)$      & $(0.47, 0.76)$    & $(0.65, 1.28)$\\
$ M_{200}/10^{14}M_{\odot} > 4$
 & $(0.45, 0.48)$ & $(0.39, 0.53)$ & $(0.36, 0.77)$\\ 
\hline

\multicolumn{4}{l}{\scriptsize
 For each bin, numbers are given
 for the noise-free ({\it left}) and noisy
  (\it right) cases.
}\\
\end{tabular}
\label{table:rawsigma}
\end{table}

\begin{table}
\caption{
Dispersion of 
$\mu_{200}=M_{\rm lens}(<\theta_{200})/M_{\rm true}(<\theta_{200})$ in
 the clean sample}
\begin{tabular}{@{}lccccc@{}} 
\hline
\hline
            &   $0.2<z<0.4$    &$0.4<z<0.6$       & $0.6<z<0.8$    \\ 
\hline
$1 < M_{200}/10^{14}M_{\odot}< 2$
&  $(0.60, 0.91)$  &    &      \\

$2 < M_{200}/10^{14}M_{\odot}< 4$
&  $(0.44, 0.63)$      & $(0.41, 0.68)$    & \\
$ M_{200}/10^{14}M_{\odot} > 4$
& $(0.32, 0.40)$ & $(0.28, 0.42)$ & $(0.25 ,0.6)$  \\ 
\hline
\multicolumn{4}{l}{\scriptsize
 For each bin, numbers are given
 for the noise-free ({\it left}) and noisy
  (\it right) cases.
}\\
\end{tabular}
\label{table:cleansigma}
\end{table}

\begin{table}
\caption{
Mean of
$\mu_{200}=M_{\rm lens}(<\theta_{200})/M_{\rm true}(<\theta_{200})$
in the clean sample}
\begin{tabular}{@{}lccccc@{}} 
\hline
\hline
            &   $0.2<z<0.4$    &$0.4<z<0.6$       & $0.6<z<0.8$    \\ 
\hline
$1 < M_{200}/10^{14}M_{\odot}< 2$

 &  $(1.02, 1.02)$  &    &      \\
$2 < M_{200}/10^{14}M_{\odot}< 4$
 &  $(1.00, 0.99)$      & $(1.11, 1.11)$    & \\
$M_{200}/10^{14}M_{\odot} > 4$
 & $(1.05, 1.04)$ & $(1.08, 1.08)$     & $(1.14, 1.12)$  \\ 
\hline
\multicolumn{4}{l}{\scriptsize
 For each bin, numbers are given
 for the noise-free ({\it left}) and noisy
  (\it right) cases.
}\\
\end{tabular} 
\label{table:cleanmean}
\end{table}

\begin{table}
\caption{Skewness of
$\mu_{200}=M_{\rm lens}(<\theta_{200})/M_{\rm true}(<\theta_{200})$
in the clean sample}
\begin{tabular}{@{}lccccc@{}} 
\hline
\hline
            &   $0.2<z<0.4$    &$0.4<z<0.6$       & $0.6<z<0.8$    \\ 
\hline
$1 < M_{200}/10^{14}M_{\odot}< 2$
 &  $(0.6, 0.16)$  &      &       \\
$2 < M_{200}/10^{14}M_{\odot}< 4$
 &  $(0.49, 0.32)$      & $(0.56, 0.13)$    &   \\
$M_{200}/10^{14}M_{\odot} > 4$
 & $(0.60, 0.11)$ & $(0.31, 0.02)$     & $(0.12, 0.07)$  \\ 
\hline
\multicolumn{4}{l}{\scriptsize
 For each bin, numbers are given 
 for the noise-free ({\it left}) and noisy
  (\it right) cases.
}\\

\end{tabular}
\label{table:cleanskewness}
\end{table}

%

\begin{table}
\caption{
Noise-free mean and dispersion 
for the ratio of the local
compensating mass 
$\Delta M(\theta_{200},1.2\theta_{200})$
to the true cluster mass $M(<\theta_{200})$}
\begin{tabular}{@{}lccccc@{}} 
\hline
\hline
            &   $0.2<z<0.4$    &$0.4<z<0.6$       & $0.6<z<0.8$    \\ 

\hline
$1 < M_{200}/10^{14}M_{\odot}< 2$
 &  $(0.25 , 0.40)$  &  &      \\
$2 < M_{200}/10^{14}M_{\odot}< 4$
 &  $(0.21 , 0.14)$      & $(0.26 , 0.16)$    &   \\
$M_{200}/10^{14}M_{\odot} > 4$
 & $(0.25 , 0.26)$ & $(0.23 , 0.29)$     & $(0.18 , 0.10)$  \\ 
\hline
\hline
\multicolumn{4}{l}{\scriptsize
 For each bin, numbers are given 
 for the mean (\it left) and the dispersion
  (\it right).
}\\

\end{tabular}
\label{table:annuluskappas}
\end{table}

\begin{table}
\caption{
Mass error dispersions 
($\sigma(\mu_{500})$, $\sigma(\mu_{1000})$, $\sigma(\mu_{1500})$)
in the clean sample for the noisy case
}
\begin{tabular}{@{}lccccc@{}} 
\hline
\hline
            &   $0.2<z<0.4$    &$0.4<z<0.6$       & $0.6<z<0.8$    \\ 
\hline
$1 < M_{200}/10^{14}M_{\odot}< 2$
 &  $(0.53, 0.39, 0.32 )$  &  &      \\
$2 < M_{200}/10^{14}M_{\odot}< 4$
 &  $(0.37, 0.28, 0.22)$      & $(0.41, 0.32, 0.28)$    &   \\
$M_{200}/10^{14}M_{\odot} > 4$
 & $(0.21, 0.17, 0.13)$ & $(0.29, 0.20, 0.18)$     & $(0.36, 0.28,
  0.25)$  \\ 
\hline
\end{tabular}
\label{table:cleansigma500}
\end{table}

\end{document}